\documentclass[sigconf]{acmart}
\acmConference[EASE 2026]{The 30th International Conference on Evaluation and Assessment in Software Engineering}{9–12 June, 2026}{Glasgow, Scotland, United Kingdom}

\setcopyright{none}
\renewcommand\footnotetextcopyrightpermission[1]{}

\usepackage{float}
\usepackage{subcaption}
\usepackage{tabularx}
\usepackage{todonotes}
\usepackage{tcolorbox}
\usepackage{amsmath}
\usepackage{balance}
\usepackage{url}
\usepackage{caption}
\usepackage{hyperref}
\usepackage{booktabs}

\title{Teaching Empathy in Software Engineering Education in the Age of Artificial Intelligence}

\author{Ronnie de Souza Santos}
\email{ronnie.desouzasantos@ucalgary.ca}
\orcid{}
\affiliation{%
  \institution{University of Calgary}
  \city{Calgary}
  \state{Alberta}
  \country{Canada}
}

\author{Cleyton Magalhães}
\email{cleyton.vanut@ufrpe.br}
\orcid{}
\affiliation{%
  \institution{Federal Rural University of Pernambuco}
  \city{Recife}
  \state{Pernambuco}
  \country{Brazil}
}

\author{Giuseppe Destefanis}
\email{g.destefanis@ucl.ac.uk}
\orcid{}
\affiliation{%
  \institution{University College London}
  \city{London}
  \country{United Kingdom}
}

\author{Mairieli Wessel}
\email{mairieli.wessel@ru.nl}
\orcid{}
\affiliation{%
  \institution{Radboud University}
  \city{Nijmegen}
  \country{The Netherlands}
}

\author{Ann Barcomb}
\email{ann.barcomb@ucalgary.ca}
\orcid{}
\affiliation{%
  \institution{University of Calgary}
  \city{Calgary}
  \state{Alberta}
  \country{Canada}
}

\author{Sherlock Licorish}
\email{sherlock.licorish@otago.ac.nz}
\orcid{}
\affiliation{%
  \institution{University of Otago}
  \city{Dunedin}
  \country{New Zealand}
}

\author{Brody-Stuart Verner}
\email{brody.stuartverner@ucalgary.ca}
\orcid{}
\affiliation{%
  \institution{University of Calgary}
  \city{Calgary}
  \state{Alberta}
  \country{Canada}
}

\author{Italo Santos}
\email{isantos3@hawaii.edu}
\orcid{}
\affiliation{%
  \institution{University of Hawaii at Manoa}
  \city{Honolulu}
  \state{Hawaii}
  \country{USA}
}

\begin{abstract}
Empathy has been discussed as a relevant human capability in software engineering, particularly in activities that require understanding users, stakeholders, and the societal implications of technological systems. This relevance becomes more pronounced in the context of artificial intelligence, where software increasingly participates in decisions that affect diverse individuals and communities. However, limited guidance exists on how empathy can be integrated into technical software engineering education in ways that connect with the development of AI enabled systems. This study investigates teaching practices that educators use to incorporate empathy into software engineering courses. Using qualitative analysis of educator reported practices, we identified five categories through which empathy is operationalized within technical coursework: societal framing of AI systems, fairness and accessibility considerations in design and evaluation, representation of diverse users, stakeholder role awareness and responsibility, and structured reflection and feedback during development processes. The findings indicate that empathy can be embedded within core development activities rather than taught as a separate topic, enabling students to reason about bias, accessibility, accountability, and the societal consequences of AI technologies. These results contribute a structured view of how empathy oriented practices can be incorporated into software engineering education to support the preparation of students who will develop AI enabled systems.
\end{abstract}

\keywords{empathy, software engineering education, artificial intelligence, responsible AI, engineering education, human-centered software development}

\begin{document}

\maketitle

\section{Introduction} \label{intro}

Large scale software development requires collaboration among developers, stakeholders, and users, which makes interpersonal capabilities relevant for successful engineering practice \cite{malinen2025soft, matturro2019systematic}. Research on software engineering skills identifies empathy as part of the set of human and communication abilities required for developers working in collaborative environments~\cite{matturro2019systematic, clear2024software}. Empathy has also been studied as an emerging human factor influencing interactions between developers and stakeholders~\cite{gunatilake2023empathy}. In software development activities such as requirements engineering and user experience design, empathy related practices help software teams better understand users’ needs and characteristics, supporting the creation of software that reflects real user expectations~\cite{ferreira2015eliciting, gunatilake2023empathy, lecca2025curious}.

The relevance of empathy becomes particularly visible in the development of artificial intelligence systems, where software increasingly participates in decisions affecting diverse groups within our society~\cite{srinivasan2022role, joo2025reimagining, schlagowski2024feeling, gates2023world, lecca2025curious}. Research on responsible AI highlights that empathy can help address the social and technical challenges involved in ensuring accountability in AI systems~\cite{srinivasan2022role, joo2025reimagining, gates2023world}. In this context, empathy supports understanding stakeholders’ intentions, needs, and contextual factors that influence how AI decisions affect individuals and communities~\cite{srinivasan2022role}. Incorporating empathy into AI design also requires recognizing the perspectives of different actors in the AI pipeline, including developers, regulators, and end users, whose expectations and responsibilities may differ~\cite{srinivasan2022role}. These observations suggest that empathy can support the development of AI systems that consider broader social and human consequences~\cite{joo2025reimagining, schlagowski2024feeling, lecca2025curious}.

These challenges highlight the importance of preparing future software engineers to recognize the human consequences of technological systems. Because of the challenges associated with understanding stakeholder perspectives, addressing accountability in AI systems, and anticipating the societal consequences of automated decisions, educational contexts have been investigated as an important environment for developing empathy among future professionals~\cite{srinivasan2022role, joo2025reimagining, gates2023world, leca2026, kelley2013teaching, bearman2015learning, stepien2006educating}. Research in education and professional training indicates that empathy can be cultivated through learning activities that encourage perspective taking, reflection, and engagement with the experiences of others~\cite{stepien2006educating, bearman2015learning, jeffrey2016empathy}. Studies in engineering and computing education also emphasize that teaching with, for, and about empathy can support social awareness, communication, and understanding of diverse user perspectives among students~\cite{kelley2013teaching, kleinrichert2024empathy}. More recent work in engineering and software engineering education further indicates that integrating empathy related activities into coursework helps students recognize the broader human and societal implications of technological systems~\cite{leca2026, sanz2025empathy}.

Several educational practices have been proposed to support the development of empathy in engineering and software engineering education, such as reflection, collaborative learning, and experiential learning, helping students develop perspective taking skills and human centered awareness in professional training~\cite{stepien2006educating, bearman2015learning, jeffrey2016empathy}. In software and computing education, approaches such as design thinking, user exploration activities, user research, persona development, and empathy mapping have been used to help students understand the needs and experiences of diverse users and consider user perspectives during system design~\cite{kelley2013teaching, kleinrichert2024empathy, sanz2025empathy, leca2026}. These pedagogical approaches also encourage teamwork, communication, and awareness of the societal implications of technical decisions, supporting the formation of engineers capable of integrating human considerations into software design~\cite{kelley2013teaching, sanz2025empathy, leca2026}.

Despite these developments, there is still limited understanding of how empathy can be systematically incorporated into the teaching of technical software engineering concepts. Existing studies discuss empathy in software engineering practice, responsible AI, and engineering education~\cite{cerqueira2023thematic, srinivasan2022role, leca2026}, but these perspectives remain fragmented when observed through an educational lens. As a result, there is limited consolidated evidence on how empathy related considerations can be integrated into technical software engineering coursework to help students understand how software systems, particularly AI systems, affect individuals and communities. To address this gap, this study investigates teaching practices that incorporate empathy into software engineering education, guided by the following question: \textbf{RQ: \textit{What teaching practices can integrate empathy into software engineering education to help students understand and address the societal impacts of AI systems?}}

This study makes three contributions to software engineering education research. First, it provides an empirically grounded characterization of how empathy can be integrated into technical coursework, showing that empathy emerges through practices that connect software development tasks with societal context, user diversity, and reflection on technological consequences. Second, it identifies and organizes a set of concrete teaching practices through which educators operationalize empathy in software engineering education, including societal framing of software problems, fairness and accessibility considerations in design and evaluation, representation of diverse users, stakeholder role awareness, and reflective critique during development processes. Third, it contributes a structured perspective on how these practices support students in developing the capacity to reason about bias, accessibility, stakeholder responsibility, and the broader societal consequences associated with AI enabled systems.

From this point, the remainder of this paper is organized as follows. Section~\ref{sec:back} presents the background on empathy in software engineering and discusses prior work that connects empathy with human centered development, stakeholder understanding, and the societal implications of AI systems. Section~\ref{sec:method} describes the research design and procedures adopted in this study, including the data collection strategy, participant characteristics, and the analytic process used to derive the categories of practices. Section~\ref{sec:findings} reports the findings of the analysis, presenting the categories of teaching practices identified by participants and illustrating how these practices integrate empathy into technical coursework. Section~\ref{sec:discussion} interprets these findings in relation to existing literature and discusses how the identified practices relate to the preparation of students for developing AI enabled systems. Finally, Section~\ref{sec:conclusion} summarizes the main contributions of the study and reflects on their relevance for software engineering education.
\section{Background}
\label{sec:back}

This section reviews prior research related to empathy in software engineering, its relevance in the development of artificial intelligence systems, and efforts to incorporate empathy into educational contexts.

\paragraph{Empathy in Software Engineering}
Empathy has increasingly been discussed as a human factor in software engineering practice \cite{cerqueira2023thematic, cerqueiraexploring}. Studies on professional competencies in software engineering identify empathy as part of the interpersonal and communication abilities required for collaboration among developers, stakeholders, and users~\cite{matturro2019systematic, clear2024software}. The literature indicates that empathy influences how developers interpret stakeholder perspectives and communicate during development activities~\cite{gunatilake2023empathy}. In requirements engineering, empathy related practices can support the elicitation of user needs by encouraging developers to consider the contexts and experiences of system users~\cite{ferreira2015eliciting}. Researchers also report increasing attention to the role of empathic understanding in developer–stakeholder interaction and human-centered development practices~\cite{cerqueira2023thematic, cerqueiraexploring, cerqueira2024empathy}. Overall, studies on this theme indicate that empathy contributes to understanding user needs and supporting human-centered software development practices.

\paragraph{Empathy in Artificial Intelligence Systems}
Empathy has also been discussed in relation to the development and evaluation of artificial intelligence systems \cite{kleinrichert2024empathy, srinivasan2022role}. Research on responsible AI indicates that empathy can support accountability in automated decision making by encouraging developers to consider the intentions, needs, and circumstances of affected stakeholders~\cite{srinivasan2022role}. Work on human-centered AI design similarly emphasizes the importance of considering human perspectives when designing and deploying intelligent technologies~\cite{joo2025reimagining}. Studies on human interaction with AI systems report that emotional and social understanding influence how users interpret and respond to intelligent systems~\cite{schlagowski2024feeling}. Additional discussions of empathy and AI technology highlight the importance of understanding how AI systems influence social contexts and human experiences~\cite{gates2023world, lecca2025curious}. In summary, studies on the intersection of these two topics suggest that empathy can support the development of AI systems that better account for human and societal considerations.

\paragraph{Teaching Empathy}
Research on empathy in education predates its discussion in software engineering and computing contexts. Early experimental work investigated empathy as a human capability that could be observed and developed through social and learning interactions~\cite{meek1957experiment}. Later, studies characterized empathy as a multidimensional construct involving perspective taking and emotional understanding~\cite{davis1990empathy}. These perspectives informed educational research exploring how instructional strategies such as reflection, experiential learning, and perspective taking exercises can support empathy development among learners~\cite{stepien2006educating, bearman2015learning, jeffrey2016empathy}. More recent studies have explored the integration of empathy-oriented activities into engineering and software engineering education, including human-centered design exercises and practices such as persona development and empathy mapping~\cite{kelley2013teaching, kleinrichert2024empathy, leca2026, sanz2025empathy}. Teaching empathy for software engineers has been acknowledged as a way to help students recognize the human contexts in which software systems operate and to consider the societal implications associated with technological systems~\cite{kelley2013teaching, kleinrichert2024empathy, leca2026, sanz2025empathy}. Collectively, these studies illustrate the progression of empathy from a concept studied in psychology and education to a topic increasingly discussed in software engineering education.
\section{Method} \label{sec:method}

This study adopted a qualitative design informed by ethnographic traditions in the social sciences and their application in software engineering research \cite{denshire2014auto, wilkinson1998focus}. Ethnography has long been used to characterize situated practices, shared meanings, and sociocultural dynamics within communities of practice \cite{wilkinson1998focus}. In software engineering, ethnographic and qualitative approaches have been employed to understand the socio technical realities surrounding everyday development work, coordination practices, and tool use \cite{sharp2010using, fincher2005making}. Although the focus of this study was pedagogical rather than industrial practice, the same orientation guided the design: to characterize the lived teaching practices and shared understandings of experienced software engineering educators across diverse institutional contexts. Our study incorporated elements of autoethnography, as each participant reflected on their own lived teaching experiences and documented individual notes capturing personal interpretations and pedagogical practices \cite{denshire2014auto}. These reflections were subsequently brought into a collective discussion setting, enabling shared sense making about empathy in software engineering education. This collective dimension aligns with methodological aspects of focus groups as spaces where meanings are articulated, negotiated, and constructed through interaction \cite{wilkinson1998focus}. The overall design is consistent with qualitative traditions in software engineering that emphasize understanding practice, context, and human factors through in depth, interaction based inquiry \cite{seaman1999qualitative}. Analytic procedures followed recommended guidance for qualitative research in software engineering, including iterative categorization, and theme development \cite{cruzes2011recommended}.

\subsection{Ethnographic Sessions Design and Scope}

Seven professors in Software Engineering took part in the study. All participants are experienced educators and researchers in the field and actively teach technical software engineering courses at undergraduate and graduate levels. A more detailed description of participant characteristics and contexts is provided in the Participants section. The main study was organized as a structured three hour collaborative session dedicated to reflecting on how empathy related practices can be incorporated into software engineering education, particularly in technically oriented courses framed by the societal impact of artificial intelligence and the fairness related characteristics of contemporary software systems. The session followed ethnographically informed procedures \cite{denshire2014auto, sharp2010using}, in that it emphasized situated reflection on lived practice, interaction based meaning construction, and iterative collective sense making rather than predefined survey instruments or hypothesis driven protocols: The session was divided into three consecutive activities.

\begin{itemize}

\item \textit{Session 1: Literature Grounding.}  
Results from a previously conducted systematic literature review on empathy in software engineering education \cite{leca2026} were presented and collectively discussed. This discussion enabled participants to relate published findings to their own situated teaching practices. Broader discussions of empathy in relation to artificial intelligence systems and fairness concerns informed the framing of this exchange.

\item \textit{Session 2: Practice Elicitation and Collective Reflection.}  
Each participant individually documented their pedagogical practices using a card based elicitation technique \cite{tchivi2025systematic}. Participants wrote on separate cards the practices they currently implement that they consider related to empathy. These practices spanned technical topics such as requirements elicitation, stakeholder analysis, accessibility considerations, fairness aware AI development, testing strategies, and project based learning. The use of a card based technique supported structured elicitation of practices while allowing participants to articulate tacit aspects of their teaching. Each participant then presented their cards to the group, followed by open discussion. This dialogic phase enabled clarification, negotiation of meanings, and comparison across institutional and cultural contexts, consistent with interaction based qualitative inquiry \cite{wilkinson1998focus}.

\item \textit{Session 3: Comparison and Categorization.}  
The practices generated during the second session were compared with those identified in the systematic literature review \cite{leca2026}. The group reflected on areas of alignment and divergence between lived teaching practice and published research. Practices were iteratively grouped and reorganized through discussion, following a dynamic card sorting process \cite{tchivi2025systematic, fincher2005making} that supported the emergence of categories grounded in participants’ own terminology and interpretations.

\end{itemize}

Although the three hour session constitutes the unit of analysis for this study, it took place within a broader five day research and academic activity focused on empathy, fairness testing, and artificial intelligence in software engineering education. The broader program included additional sessions dedicated to fairness testing and intersectionality in AI development, as well as a panel discussion on the impact of AI in education. These components are not directly analyzed in this paper; however, the extended engagement with fairness, intersectionality, and AI informed participants’ reflections and contributed to the development and refinement of the categories and themes that emerged from the card sorting process and subsequent analysis.

\subsection{Participants}

Seven professors in Software Engineering participated in the session. All had more than five years of teaching experience in software engineering and more than ten years of research experience in the field. All currently hold academic positions and are actively involved in curriculum design, course coordination, supervision of graduate students, and research projects. All participants teach at both undergraduate and graduate levels. Their teaching portfolios include core software engineering courses such as requirements engineering, software architecture, software testing, and project based capstone courses. Several have experience teaching large enrollment foundational courses as well as smaller seminar style graduate courses. All have taught courses requiring hands on development activities, including coding assignments, collaborative projects, and full software deployment. The group represents diverse research expertise. All participants conduct research on human aspects of software engineering. Three have conducted sustained research on software fairness and bias in software systems. Five investigate societal impacts of artificial intelligence. Two conduct research on conversational agents and chatbots. Two work with mining software repositories and data driven software engineering. All seven are active contributors to software engineering education research. Participants teach in different national and institutional contexts, including Canada, the United States, Brazil, the United Kingdom, the Netherlands, and New Zealand. These contexts represent several regions with distinct educational systems, cultural norms, regulatory environments, and societal debates concerning AI and fairness. This geographic diversity enabled reflection on how empathy oriented practices are shaped by local curricular structures, student demographics, institutional priorities, and cultural expectations regarding equity and inclusion.

\subsection{Data Collection and Analysis}

\begin{figure}[h]
\centering
\includegraphics[width=0.8\linewidth]{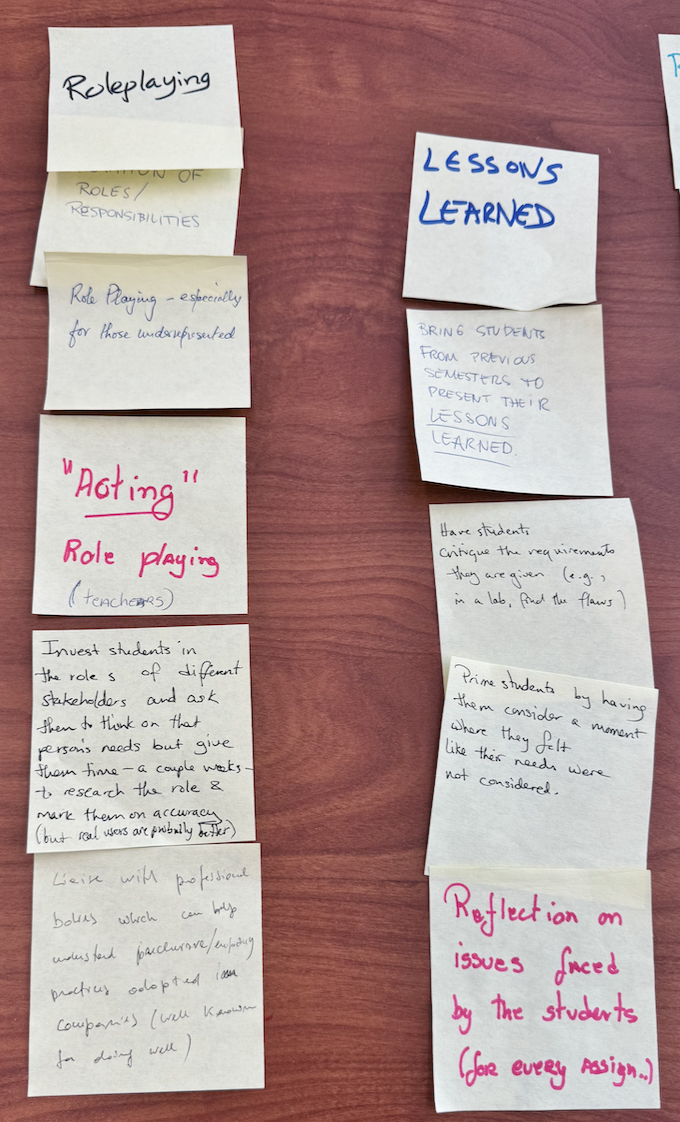}
\caption{Card Sorting Process}
\label{fig:card}
\end{figure}

Data consisted of three sources: (1) individual written notes produced by participants during reflective moments of the session, (2) the set of practice cards generated in the second phase, and (3) collective notes documenting key points from the group discussion. Figure \ref{fig:card} is a real example of cards being grouped during the process. Data analysis followed an iterative thematic process consistent with recommended practices for qualitative analysis in software engineering \cite{cruzes2011recommended}. First, the cards and notes were considered as low level codes, assigned to individual practices, capturing specific pedagogical actions and strategies. Second, related low level codes were grouped into higher level themes representing broader pedagogical intentions. Third, these themes were organized into overarching categories representing families of empathy related practices in software engineering education. The categorization was supported by the dynamic card sorting process \cite{fincher2005making}. 

Practices were iteratively grouped, reorganized, merged, and refined during discussion until a stable structure was reached across multiple interactions. Throughout the analysis, attention was given to how participants articulated their rationale, contextual constraints, and intended outcomes. The interactive setting enabled observation of how shared understandings about empathy, fairness, and AI in software systems were constructed and refined through dialogue \cite{wilkinson1998focus}. The resulting categories and associated practices are presented in the Results section. For illustration, the analytic process can be exemplified with one practice. A low-level practice verbatim from the cards, such as:

\begin{quote}
\textit{“Using AI tools to simulate how someone of an underrepresented group would respond.”}
\end{quote}

was initially treated as an individual code capturing a concrete pedagogical action. During iterative comparison with related practices, it was grouped under the theme \textit{Bias Detection and Accessibility Awareness}, which encompassed practices concerned with identifying overlooked user needs and recognizing structural bias in system design. At a higher level of abstraction, this theme was subsequently organized under the final category \textit{Fairness, Bias, and Accessibility in Design and Evaluation}. This example illustrates how a specific instructional action progressed from a low-level code to a theme and ultimately to a broader final category through iterative grouping and collective discussion.

\subsection{Threats to Validity}

This study is qualitative and ethnographically informed; therefore, threats to validity are considered in terms of credibility, transferability, and dependability rather than statistical generalization. \textit{Researcher positionality and interpretive bias.}  Ethnographic and autoethnographic approaches are inherently interpretive, and researchers’ prior experiences and theoretical orientations may influence how practices are articulated and categorized. In this study, all participants are experienced researchers in software engineering, and several work directly on empathy, fairness, and AI. This shared background may introduce confirmation bias. We addressed this risk by incorporating multiple perspectives across seven participants from different countries and institutional contexts, by structuring the session to include both individual elicitation and collective discussion, and by iteratively revisiting emerging categories until they stabilized across interactions. \textit{Reactivity and group dynamics.}  Focus group settings may introduce social desirability effects or dominance dynamics, where some voices shape the discussion more strongly than others. To mitigate this threat, participants first documented practices individually using a card based elicitation technique before engaging in group discussion. This sequence reduced immediate convergence and ensured that each participant’s perspective was explicitly represented in the data prior to collective negotiation. \textit{Limited immersion.} Although informed by ethnographic principles, the focal unit of analysis was a structured three hour session rather than prolonged field immersion. As such, the study captures articulated teaching practices rather than direct classroom observation. We acknowledge this limitation and frame the study as ethnographically informed rather than full immersion ethnography. \textit{Card sorting limitations.} Card sorting may lead to surface level categorization or simplification of nuanced practices. To address this, categories were not fixed a priori. Instead, practices were iteratively grouped and reorganized through discussion, and original card wording was preserved to maintain traceability between low level practices and higher level themes. \textit{Transferability.}  Our findings emerge from seven experienced professors working in specific educational contexts. We do not claim statistical generalization. Instead, consistent with qualitative research traditions, we aim for analytical transferability. By providing detailed descriptions of participants, contexts, procedures, and analytic steps, we enable readers to assess the extent to which the findings may be transferable to other software engineering educational settings.
\section{Findings} \label{sec:findings}

Our study resulted in five categories that organize the practices participants identified for teaching empathy within technical software engineering education. Here, these practices are framed within the context of the AI era, particularly considering how students can learn about empathy into the development of software systems, especially AI-enabled systems. Table \ref{tab:hierarchical-practices} presents the complete hierarchical analysis, including final categories, themes, and verbatim practices.

\paragraph{AI Systems and Societal Reality Framing} This group of practices situates technical development work within broader societal contexts. Educators adopting this approach design assignments that explicitly connect software to real-world challenges and affected communities. For instance, some participants require students to focus development efforts on “developing / modelling for underrepresented groups (water, pollution, poverty)” or to “bring evidence of their ‘project problem’ (e.g., news, papers).” Others require that students “come with ideas / fixes / etc. considering sustainability challenges” or analyze “news items (focusing on security flaws, impacts on users, inequality, etc.) on software.” In practice, this means that instructors embed contextual analysis, societal grounding, and sustainability reasoning directly into requirements engineering and project framing. Within this category, empathy is operationalized as contextual awareness. When considering aspects related to AI-powered applications, these practices encourage students to recognize that AI systems are not neutral artifacts but interventions in complex social environments. Empathy becomes the capacity to anticipate societal impact, understand structural inequality, and incorporate real-world consequences into technical decision making.

\paragraph{AI Fairness, Bias, and Accessibility in Design and Evaluation}

This group of practices integrates empathy directly into design and evaluation activities. Educators adopting this approach embed accessibility and fairness considerations into technical assignments and grading criteria. For example, some participants include “design considerations – considering differently abled people” as an explicit requirement. Others assign part of grading on “style” to someone “from the community with accessibility requirements,” rotating the evaluator so students “cannot design for just one person.” Additional practices require students to “infer requirements from reviews (e.g., app reviews)” to identify overlooked accessibility needs, or to “critically analyze teaching / learning material, taking into account biases and stereotypes.” In some cases, instructors also incorporate tools by “using AI tools to simulate how someone of an underrepresented group would respond.” In practice, educators embed fairness and accessibility into core technical evaluation processes rather than treating them as peripheral topics. Within this category, empathy is operationalized in AI-related contexts as fairness-oriented technical competence. When considering AI-powered applications, students are encouraged to recognize that algorithmic outputs and automated decisions may affect users differently. Empathy becomes the ability to detect bias, anticipate exclusion, and integrate accessibility and fairness considerations into system design, testing, and evaluation processes.

\paragraph{Representative Users and Inclusion in AI System Development}

This group of practices centers on incorporating diverse user perspectives into technical coursework and learning environments. By adopting this approach, educators create structured opportunities for direct engagement with user experiences. For instance, some participants “bring a representative of ‘the user’ to the lecture (e.g., blind person)” or provide “modified requirements but get real people to present them as their needs and explanation as to why they matter.” Others intentionally diversify examples used “in tests, in slides,” ensuring gender and racial diversity, and draw on “different hobbies, themes, experiences” when constructing software scenarios. At the same time, participants stress the importance to “be careful to avoid tokenizing people in the class.” In practice, instructors deliberately diversify representation and embed inclusion into requirements discussions and technical examples. Within this category, empathy is operationalized in AI development as inclusive representation. When designing AI-powered systems, students are invited to reflect that training data, user models, and system assumptions may privilege certain groups while excluding others. Empathy becomes the practice of incorporating diverse user perspectives into system development processes and ensuring that representation is not superficial but meaningfully integrated.

\paragraph{Stakeholder Role Awareness and Responsibility in AI Systems}

This group of practices focuses on structured perspective taking and explicit articulation of roles and responsibilities within development processes. Educators adopting this approach define “roles / responsibilities” in assignments and implement activities such as “role playing – especially for those underrepresented” and extended exercises where students “invest … in the roles of different stakeholders” and are assessed on the accuracy of their representation. Additional practices include liaising “with professional bodies which can help understand practices adopted from companies” and ensuring “diversity in team formation – starting with gender and course background.” In practice, instructors simulate stakeholder dynamics and distribute accountability across technical roles. Within this category, empathy is operationalized in AI contexts as responsibility-aware perspective taking. When working with AI systems, students need to understand that decision making, model selection, deployment, and monitoring involve multiple stakeholders with different needs and risks. Empathy becomes the capacity to anticipate how technical decisions affect various actors and to recognize distributed accountability across the AI development lifecycle.

\paragraph{Critical Reflection, Feedback, and Iterative Awareness in AI Education}

This group of practices emphasizes structured reflection and iterative critique within technical coursework. These practices consider using reflection prompts such as “reflection on issues faced by the students (for every assignment)” and ask students to consider “a moment where they felt like their needs were not considered.” Students are also required to “critique the requirements they are given (e.g., in a lab, find the flaws)” and engage in “peer evaluation – especially for teams / individuals with different skill levels.” Some participants invite “students from previous semesters to present their lessons learned.” In practice, instructors integrate reflection and feedback mechanisms into assignments and team activities. Within this category, empathy is operationalized in AI-related development as iterative awareness of unintended consequences. When building AI-powered applications, students are can explore overlooked needs, hidden assumptions, and potential harms. Empathy becomes an ongoing evaluative process, supporting continuous refinement of technical artifacts and heightened sensitivity to exclusion and bias.

\begin{table*}[t]\scriptsize
\centering
\caption{Hierarchical organization of empathy-related teaching practices}
\label{tab:hierarchical-practices}
\footnotesize
\begin{tabular}{p{2cm} p{2.5cm} p{10cm}}
\toprule
\textbf{Final Category} & \textbf{Theme} & \textbf{Low-Level Practice (Verbatim from Cards)} \\
\midrule

AI Systems and Societal Reality Framing &
Real-World Societal Context Integration &
Focus area of dev – developing / modelling for underrepresented groups (water, pollution, poverty). \\

 &
 &
In terms of requirements, come up with ideas that could be applied in real contexts (small projects). \\

 &
 &
Bring evidence of their “project problem” (e.g., news, papers). \\

\cmidrule(lr){2-3}

 &
Sustainability and Structural Impact Awareness &
Have to come with ideas / fixes / etc. considering sustainability challenges. \\

 &
 &
News items (focusing on security flaws, impacts on users, inequality, etc.) on software – give them the task of bringing one news item to class. \\

\cmidrule(lr){2-3}

 &
Economic and Industry Framing of Responsibility &
Make the business case for EDIA, not just the moral case. Some will only care about \$. \\

 &
 &
I used to have a panel discussion of people working in industry talking about how they use the techniques in class in their work. \\

\midrule

AI Fairness, Bias, and Accessibility in Design and Evaluation &
Accessibility-Oriented Design &
Design considerations – considering differently abled people. \\

 &
 &
Give a subset of grading on “style” to be marked by someone from the community with accessibility requirements – pick a different person each year so they cannot design for just one person. \\

\cmidrule(lr){2-3}

 &
Bias Detection and Critical Analysis &
Show how to critically analyze teaching / learning material, taking into account biases and stereotypes. \\

 &
 &
Have students infer requirements from reviews (e.g., app reviews) and see if they pick out accessibility etc., and grade them on it. When they see it is unfair, see they were too. \\

\cmidrule(lr){2-3}

 &
AI-Mediated Simulation of Marginalized Perspectives &
Using AI tools to simulate how someone of an underrepresented group would respond. \\

\midrule

Representative Users and Inclusion in AI System Development &
Direct User Representation &
Bring a representative of “the user” to the lecture (e.g., blind person). \\

 &
 &
Give modified requirements but get real people to present them as their needs and explanation as to why they matter. \\

\cmidrule(lr){2-3}

 &
Inclusion in Classroom Design Feedback &
Call underrepresented students into the loop and let them explain what aspects should be improved / modified. \\

\cmidrule(lr){2-3}

 &
Diverse Examples and Anti-Tokenization &
Make sure your examples (in tests, in slides) are gender and racially diverse (I literally roll a dice and take first names from my contacts). NB: Must have diverse context. \\

 &
 &
Make sure your examples of software draw on different hobbies, themes, experiences and tell about what you did and why. \\

 &
 &
Be careful to avoid tokenizing people in the class and bring them to teach the group to recognize their needs and humanity. \\

\midrule

Stakeholder Role Awareness and Responsibility in AI Systems &
Explicit Role Definition and Accountability &
Definition of roles / responsibilities. \\

\cmidrule(lr){2-3}

 &
Role-Based Perspective Taking &
Role playing – especially for those underrepresented. \\

 &
 &
“Acting” – role playing (teachers). \\

 &
 &
Invest students in the roles of different stakeholders and ask them to think on that person’s needs but give them time – a couple weeks – to research the role and mark them on accuracy (but real users are probably better). \\

\cmidrule(lr){2-3}

 &
Professional and Team Responsibility Structures &
Liaise with professional bodies which can help understand practices adopted from companies (well known for doing well). \\

 &
 &
Diversity in team formation – starting with gender and course background. \\

\midrule

Critical Reflection, Feedback, and Iterative Awareness in AI Education &
Structured Reflection Practices &
Reflection on issues faced by the students (for every assignment). \\

 &
 &
Prime students by having them consider a moment where they felt like their needs were not considered. \\

\cmidrule(lr){2-3}

 &
Critical Review and Error Identification &
Have students critique the requirements they are given (e.g., in a lab, find the flaws). \\

\cmidrule(lr){2-3}

 &
Peer Learning and Iterative Feedback &
Peer evaluation – especially for teams / individuals with different skill levels. \\

 &
 &
Bring students from previous semesters to present their lessons learned. \\

\bottomrule
\end{tabular}
\end{table*}
\section{Discussion}
\label{sec:discussion}

This section discusses our findings in relation to the research question guiding this study. We first synthesize our findings by directly answering the research question. We then relate our findings to existing literature on empathy in software engineering education and discuss implications for research and educational practice.

\subsection{Answering the Research Question}

Our research question asked: \textit{What teaching practices can integrate empathy into software engineering education to help students understand and address the societal impacts of AI systems?} Our findings suggest that empathy can be integrated into software engineering education by embedding social and human considerations within core technical learning activities. Rather than being taught as a separate topic, empathy appears as a capability developed through practices that connect software development tasks with societal context, user diversity, stakeholder perspectives, and reflection on technological consequences. In the context of the AI era and the development of AI-enabled systems, these practices help students understand how algorithmic decisions interact with social environments, institutional contexts, and diverse user experiences. Across the categories identified in our analysis, such practices encourage students to situate AI systems within real-world contexts, recognize how technical decisions affect different users and communities, consider the perspectives and responsibilities of multiple stakeholders, and reflect on the potential impacts and limitations of technological solutions.  Through these practices, empathy supports students in reasoning about how software systems interact with social environments and recognizing that AI-enabled systems are not neutral artifacts but sociotechnical interventions with consequences for individuals and communities.

\subsection{Comparison with the Literature} 

Prior work connects empathy in software engineering with stakeholder understanding, communication, and human-centered development practices \cite{matturro2019systematic, clear2024software, gunatilake2023empathy, ferreira2015eliciting}. Our findings align with this perspective insofar as several practices identified by participants emphasize understanding users and situating development decisions within real-world contexts. In particular, the categories \textit{AI systems and societal reality framing} and \textit{representative users and inclusion in AI system development} show how educators can connect technical coursework with the lived experiences of individuals and communities. These practices reinforce the view that software engineering activities benefit from considering social contexts and user perspectives \cite{cerqueira2023thematic, cerqueiraexploring, cerqueira2024empathy}. Our findings also align with discussions explicitly connecting empathy to the societal implications of artificial intelligence systems. The literature on this topic emphasizes the need to account for stakeholder perspectives, accountability, and social consequences when designing AI technologies \cite{srinivasan2022role, joo2025reimagining}. In our findings, the category \textit{AI fairness, bias, and accessibility in design and evaluation} reflects these concerns by incorporating accessibility analysis, bias identification, and fairness considerations into technical assignments and evaluation activities. This suggests that educators increasingly frame empathy in relation to the broader impacts that AI systems may have on diverse groups of users \cite{schlagowski2024feeling, gates2023world, lecca2025curious, devathasan2025empathy}.

At the same time, our findings extend existing educational approaches that introduce empathy through user-centered or design-oriented activities \cite{kelley2013teaching, kleinrichert2024empathy, leca2026, sanz2025empathy}. While previous studies often present empathy-related exercises as individual pedagogical techniques, our findings indicate that educators should integrate empathy across multiple stages of technical coursework. The practices identified in our study connect contextual analysis, fairness-aware design, stakeholder perspective exploration, and reflective critique with core software engineering activities. Additionally, our study show how empathy can be integrated into practices that simulate professional responsibilities in software and AI development. The categories \textit{stakeholder role awareness and responsibility in AI systems} and \textit{critical reflection, feedback, and iterative awareness in AI education} illustrate how educators can encourage students to analyze responsibilities across different actors and to reflect on the consequences of technical decisions. These practices connect empathic reasoning with accountability in development processes.

Overall, our findings contribute a structured view of how empathy can be incorporated into technical software engineering education within the context of AI development. The five categories identified in our analysis illustrate how educators operationalize empathy through societal framing of problems, fairness-aware design and evaluation, representation of diverse users, stakeholder role awareness, and iterative reflection during technical coursework.

\subsection{Implications}

\textbf{For researchers}, our findings indicate that research on empathy in software engineering education would benefit from moving beyond conceptual discussions toward systematic investigation of how empathy is embedded in technical coursework. While prior literature frequently connects empathy with user centered design and stakeholder communication, the practices identified in this study indicate that empathy can also be integrated through activities that address societal framing of problems, fairness aware design and evaluation, representation of diverse users, stakeholder responsibility, and structured reflection on technological consequences. Future research could investigate how these forms of integration appear across different types of software engineering courses, such as requirements engineering, software design, testing, and artificial intelligence related subjects. Additional research could also analyze how empathy oriented practices influence students’ reasoning about bias, accessibility, and accountability in AI enabled systems. Longitudinal studies would be particularly useful to characterize whether exposure to these practices contributes to changes in students’ awareness of societal impacts and their ability to reason about the consequences of technical decisions. Comparative studies across institutions and curricular structures could further contribute to understanding how empathy oriented practices can be incorporated into established software engineering curricula without displacing core technical learning outcomes.

\textbf{For educators}, our findings indicate that empathy can be incorporated into software engineering education without requiring separate courses dedicated to social topics. Instead, the practices we identified and explored suggest that empathy becomes visible when social and human considerations are integrated into the structure of technical assignments and project activities. Educators can situate technical work within real world contexts by asking students to analyze societal problems, consult evidence from news or research sources, and consider how software systems may affect different communities. Design and evaluation activities can incorporate accessibility and fairness considerations so that students learn to identify potential bias and exclusion during system development. Opportunities for interaction with diverse user perspectives, such as engagement with representative users or diversified examples in requirements scenarios, can help students recognize variation in user needs and experiences. Structured role based activities can further encourage students to consider the responsibilities of different stakeholders involved in AI development processes. Finally, reflection and feedback mechanisms embedded within assignments can support continuous reconsideration of assumptions and consequences during development activities. These practices indicate that empathy oriented learning can be incorporated into the existing structure of software engineering education while maintaining the technical focus of the discipline. In the context of the AI era, this integration becomes particularly relevant as students prepare to work on systems whose automated decisions interact with complex social environments. Through the practices identified in this study, students can develop the capacity to consider how data, models, and design choices may affect different groups of users, anticipate potential harms or exclusions, and reflect on the responsibilities associated with developing AI enabled systems. In this sense, empathy oriented practices support the preparation of future software engineers who are able to reason about both the technical and societal implications of AI technologies.
\section{Conclusions and Future Work}
\label{sec:conclusion}

This study investigated how empathy can be integrated into software engineering education to support students in understanding and addressing the societal impacts of AI enabled systems. Based on educators’ reported practices, the findings indicate that empathy can be incorporated into technical coursework through activities that connect software development tasks with societal context, user diversity, stakeholder perspectives, and reflection on technological consequences. Rather than appearing as a separate instructional topic, empathy emerges through pedagogical practices embedded in core development activities. The analysis identified five categories of practices through which educators operationalize empathy in the classroom: societal framing of AI systems, fairness and accessibility considerations in design and evaluation, representation of diverse users, stakeholder role awareness and responsibility, and structured reflection and feedback during development processes. Together, these practices illustrate how empathy can be integrated into the technical structure of software engineering education while supporting students in reasoning about bias, accessibility, responsibility, and the broader societal consequences associated with AI systems.

Our future work will extend this research through empirical investigation of how these practices influence students’ learning experiences during the development of AI powered systems. In particular, we plan to design classroom interventions that operationalize the categories identified in this study and evaluate them through experimentation in software engineering courses. These interventions will allow us to analyze how the practices affect students’ awareness of societal impacts, their reasoning about fairness and accessibility, and their capacity to reflect on the consequences of algorithmic decisions. In addition, we plan to investigate how empathy oriented practices relate to professional competencies expected in entry level software engineering roles, particularly in projects involving AI technologies. By analyzing the relationship between these practices and skills requested by industry, we aim to better understand how educational strategies that integrate empathy may contribute to preparing software engineers to work responsibly with AI enabled systems.

\balance
\bibliographystyle{ACM-Reference-Format}
\bibliography{biblio}

@inproceedings{cerqueira2023thematic,
  title={A thematic synthesis on empathy in software engineering based on the practitioners' perspective},
  author={Cerqueira, Lidiany and Freire, S{\'a}vio and Bastos, Jo{\~a}o and Sp{\'\i}nola, Rodrigo and Mendon{\c{c}}a, Manoel and Santos, Jos{\'e}},
  booktitle={Proceedings of the XXXVII Brazilian Symposium on Software Engineering},
  pages={332--341},
  year={2023}
}

@article{gunatilake2023empathy,
  title={Empathy models and software engineering—A preliminary analysis and taxonomy},
  author={Gunatilake, Hashini and Grundy, John and Mueller, Ingo and Hoda, Rashina},
  journal={Journal of Systems and Software},
  volume={203},
  pages={111747},
  year={2023},
  publisher={Elsevier}
}

@article{cerqueiraexploring,
  title={Exploring Empathy in Software Engineering: Insights from a Grey Literature Analysis of Practitioners’ Perspectives},
  author={Cerqueira, Lidiany and Bastos, Jo{\~a}o Pedro and Neves, Danilo and Carneiro, Glauco and Spinola, Rodrigo and Freire, S{\'a}vio and Santos, Jose and Mendon{\c{c}}a, Manoel},
  journal={ACM Transactions on Software Engineering and Methodology},
  publisher={ACM New York, NY}
}

@article{cerqueira2024empathy,
  title={Empathy and its effects on software practitioners’ well-being and mental health},
  author={Cerqueira, Lidiany and Freire, S{\'a}vio and Neves, Danilo Ferreira and Bastos, Jo{\~a}o Pedro Silva and Santana, Beatriz and Sp{\'\i}nola, Rodrigo and Mendon{\c{c}}a, Manoel and Santos, Jos{\'e} Amancio Macedo},
  journal={IEEE Software},
  volume={41},
  number={4},
  pages={95--104},
  year={2024},
  publisher={IEEE}
}

@article{clear2024software,
  title={Software Developers and Collective Empathy—Can They Be Disposed to Care?},
  author={Clear, Tony},
  journal={A Transformative Model for Open Access},
  pages={6},
  year={2024}
}

@inproceedings{lecca2025curious,
  title={Curious, critical thinker, empathetic, and ethically responsible: Essential soft skills for data scientists in software engineering},
  author={Le{\c{c}}a, Matheus De Morais and Santos, Ronnie De Souza},
  booktitle={2025 IEEE/ACM 47th International Conference on Software Engineering: Software Engineering in Society (ICSE-SEIS)},
  pages={151--162},
  year={2025},
  organization={IEEE}
}

@article{devathasan2025empathy,
  title={Empathy, self-determination and motivation: moderating diversity for enhanced performance in software development teams},
  author={Devathasan, Kezia and Arony, Nowshin Nawar and Murphy-Hill, Emerson and Damian, Daniela},
  journal={Empirical Software Engineering},
  volume={30},
  number={3},
  pages={82},
  year={2025},
  publisher={Springer}
}

@inproceedings{ferreira2015eliciting,
  title={Eliciting requirements using personas and empathy map to enhance the user experience},
  author={Ferreira, Bruna and Conte, Tayana and Barbosa, Simone Diniz Junqueira},
  booktitle={2015 29th Brazilian symposium on software engineering},
  pages={80--89},
  year={2015},
  organization={IEEE}
}

@article{matturro2019systematic,
  title={A systematic mapping study on soft skills in software engineering.},
  author={Matturro, Gerardo and Raschetti, Florencia and Font{\'a}n, Carina},
  journal={J. Univers. Comput. Sci.},
  volume={25},
  number={1},
  pages={16--41},
  year={2019}
}

@inproceedings{malinen2025soft,
  title={Soft Skills in Software Engineering: Insights from the Trenches},
  author={Malinen, Sanna and Galster, Matthias and Mitrovic, Antonija and Iyer, Sreedevi Sankara and Peirisand, Pasan and Clarke, April},
  booktitle={2025 IEEE/ACM 47th International Conference on Software Engineering: Software Engineering in Practice (ICSE-SEIP)},
  pages={296--306},
  year={2025},
  organization={IEEE}
}

@inproceedings{cruzes2011recommended,
  title={Recommended steps for thematic synthesis in software engineering},
  author={Cruzes, Daniela S and Dyba, Tore},
  booktitle={2011 international symposium on empirical software engineering and measurement},
  pages={275--284},
  year={2011},
  organization={IEEE}
}

@article{stepien2006educating,
  title={Educating for empathy: a review},
  author={Stepien, Kathy A and Baernstein, Amy},
  journal={Journal of general internal medicine},
  volume={21},
  number={5},
  pages={524--530},
  year={2006},
  publisher={Springer}
}

@article{jeffrey2016empathy,
  title={Empathy--can it be taught?},
  author={Jeffrey, David and Downie, Robert},
  journal={Journal of the Royal College of Physicians of Edinburgh},
  volume={46},
  number={2},
  pages={107--112},
  year={2016},
  publisher={SAGE Publications Sage UK: London, England}
}

@article{bearman2015learning,
  title={Learning empathy through simulation: a systematic literature review},
  author={Bearman, Margaret and Palermo, Claire and Allen, Louise M and Williams, Brett},
  journal={Simulation in healthcare},
  volume={10},
  number={5},
  pages={308--319},
  year={2015},
  publisher={LWW}
}

@article{davis1990empathy,
  title={What is empathy, and can empathy be taught?},
  author={Davis, Carol M},
  journal={Physical therapy},
  volume={70},
  number={11},
  pages={707--711},
  year={1990},
  publisher={Oxford University Press}
}

@article{kelley2013teaching,
  title={Teaching empathy and other compassion-based communication skills},
  author={Kelley, Kevin J and Kelley, Mary F},
  journal={Journal for nurses in professional development},
  volume={29},
  number={6},
  pages={321--324},
  year={2013},
  publisher={LWW}
}

@article{meek1957experiment,
  title={An experiment in teaching empathy},
  author={Meek, Clinton R},
  journal={The Journal of Educational Sociology},
  volume={31},
  number={2},
  pages={107--110},
  year={1957},
  publisher={JSTOR}
}

@article{gates2023world,
  title={A world beyond self: Empathy and pedagogy during times of global crisis},
  author={Gates, Eliza and Curwood, Jen Scott},
  journal={The Australian Journal of Language and Literacy},
  volume={46},
  number={2},
  pages={195--209},
  year={2023},
  publisher={Springer}
}

@article{srinivasan2022role,
  title={The role of empathy for artificial intelligence accountability},
  author={Srinivasan, Ramya and Gonz{\'a}lez, Beatriz San Miguel},
  journal={Journal of Responsible Technology},
  volume={9},
  pages={100021},
  year={2022},
  publisher={Elsevier}
}

@article{schlagowski2024feeling,
  title={The feeling of being classified: raising empathy and awareness for AI bias through perspective-taking in VR},
  author={Schlagowski, Ruben and Volanti, Maurizio and Weitz, Katharina and Mertes, Silvan and Kuch, Johanna and Andr{\'e}, Elisabeth},
  journal={Frontiers in Virtual Reality},
  volume={5},
  pages={1340250},
  year={2024},
  publisher={Frontiers Media SA}
}

@article{kleinrichert2024empathy,
  title={Empathy: an ethical consideration of AI \& others in the workplace},
  author={Kleinrichert, Denise},
  journal={AI \& SOCIETY},
  volume={39},
  number={6},
  pages={2743--2757},
  year={2024},
  publisher={Springer}
}

@article{sanz2025empathy,
  title={Empathy, bias, and data responsibility: evaluating AI chatbots for gender-based violence support},
  author={Sanz Urquijo, Borja and L{\'o}pez Belloso, Mar{\'\i}a and Izaguirre-Choperena, Ainhoa},
  journal={Frontiers in Political Science},
  volume={7},
  pages={1631881},
  year={2025},
  publisher={Frontiers Media SA}
}

@inproceedings{joo2025reimagining,
  title={Reimagining Interaction through Empathy and Ethics: AI with a Human-Centered Approach},
  author={Joo, Nazim Hussain and Vij, Bhoomi},
  booktitle={2025 International Conference on Responsible, Generative and Explainable AI (ResGenXAI)},
  pages={1--7},
  year={2025},
  organization={IEEE}
}

@inproceedings{leca2026,
  title={Empathy in Software Engineering Education: Evidence, Practices, and Opportunities},
  author={Leca, Matheus de Morais and Johnston, Kim and de Souza Santos, Ronnie},
  booktitle={19th International Conference on Cooperative and Human Aspects of Software Engineering (CHASE’26)},
  pages={1--11},
  year={2026},
  organization={ACM}
}

@article{denshire2014auto,
  title={On auto-ethnography},
  author={Denshire, Sally},
  journal={Current Sociology},
  volume={62},
  number={6},
  pages={831--850},
  year={2014},
  publisher={Sage Publications Sage UK: London, England}
}

@inproceedings{sharp2010using,
  title={Using ethnographic methods in software engineering research},
  author={Sharp, Helen and deSouza, Cleidson and Dittrich, Yvonne},
  booktitle={Proceedings of the 32nd ACM/IEEE International Conference on Software Engineering-Volume 2},
  pages={491--492},
  year={2010}
}

@article{wilkinson1998focus,
  title={Focus group methodology: a review},
  author={Wilkinson, Sue},
  journal={International journal of social research methodology},
  volume={1},
  number={3},
  pages={181--203},
  year={1998},
  publisher={Taylor \& Francis}
}

@article{fincher2005making,
  title={Making sense of card sorting data},
  author={Fincher, Sally and Tenenberg, Josh},
  journal={Expert Systems},
  volume={22},
  number={3},
  pages={89--93},
  year={2005},
  publisher={Wiley Online Library}
}

@article{tchivi2025systematic,
  title={A systematic review of the comparison of different types of card sorting},
  author={Tchivi, Elinda and Sharma, Bibhya and Paea, Sione},
  journal={IEEE Access},
  year={2025},
  publisher={IEEE}
}

@article{seaman1999qualitative,
  title={Qualitative methods in empirical studies of software engineering},
  author={Seaman, Carolyn B.},
  journal={IEEE Transactions on software engineering},
  volume={25},
  number={4},
  pages={557--572},
  year={1999},
  publisher={IEEE}
}

\end{document}